\documentclass[aps,prd,preprint,superscriptaddress,nofootinbib]{revtex4-2}
\usepackage{graphicx}
\usepackage{dcolumn}
\usepackage{bm}
\usepackage{multirow}
\usepackage{color}%
\usepackage[english]{babel}
\usepackage{amsmath,amssymb}
\usepackage{amsfonts}
\usepackage{mathrsfs}
\usepackage{epsfig}
\usepackage{tabularx}
\usepackage{array}
\usepackage{slashed}
\usepackage{breqn}
\usepackage{braket}
\usepackage{mathtools}

\usepackage{floatrow}   
\usepackage{tabularray}
\usepackage{xcolor}
\usepackage{makecell}
\usepackage{hyperref}
\usepackage{siunitx}
\usepackage{booktabs}
\allowdisplaybreaks

\graphicspath{{./figs/}}

\DeclarePairedDelimiter\autobracket{(}{)}
\newcommand{\br}[1]{\autobracket*{#1}}

\newcommand{\mi}{\mathrm{i}}

\newcommand{\checked}[2][0.5cm]{%
	\noindent\parbox[t]{#1}{\raggedright\ding{51}}\parbox[t]{\the\dimexpr\linewidth-#1}{#2}%
}

\newcolumntype{Y}{>{\centering\arraybackslash}X} 
\newcolumntype{S}{>{\centering\arraybackslash}S[table-format=-1.2]} 

\def\s2tw{{\rm sin ^2 \theta_{W}}}

\makeatletter
\def\slashchar#1{{\mathpalette\c@ncel{#1}}} 
\makeatother

\makeatletter
\begin{document}


\title{Explaining Belle Data on $B\to K^{(*)}\nu\bar{\nu}$ Decays via Dark $Z$ Resonances}

\newcommand{\metu}{Department of Physics, Middle East Technical University, Ankara 06800, Turkey.}
\newcommand{\au}{Department of Engineering Physics, Ankara	University, TR06100 Ankara, Turkey.}

\author{T.M.~Aliev}  \affiliation{ \metu }
\author{A. Elpe}  \affiliation{ \metu }
\author{L. Selbuz} \affiliation{ \au }
\author{I.~Turan}  \affiliation{ \metu }


\date{\today}

\begin{abstract}
The Belle II Collaboration reported the first measurement on ${\rm Br}(B^+\rightarrow K^+\nu\bar{\nu})$, which lies 2.7$\sigma$ away from the Standard Model expectation. This result may be manifestation of new physics beyond the Standard Model. In present work, motivated by the Belle II measurement, we investigate the effect of a dark photon/dark $Z$ on the rare $B$ meson decay $B^+\rightarrow K^+\nu\bar{\nu}$ and show that the ${\rm Br}(B^+\rightarrow K^+\nu\bar{\nu})$ excess from Belle II over the Standard Model expectation explained by the appearance of the dark $Z$ especially in  resonances. We also derive constraints on various parameters of the two-Higgs-doublet model extended with a dark Abelian gauge group, in light of the measurement of the ${\rm Br}(B^+\rightarrow K^+\nu\bar{\nu})$ by the Belle-II Collaboration as well as the upper bound on ${\rm Br}(B^0\rightarrow K^{*0}\nu\bar{\nu})$ decay, set by the Belle data. Our results indicate that there exists a common region where both experimental results are satisfied for a dark $Z$ mass around 4.5 GeV with suitable values of the other parameters.

\end{abstract}


\maketitle
\newpage
 
 
\section{Introduction}
\label{intro}

The observation of the Higgs boson at the CERN Large Hadron Collider \cite{ATLAS:2012yve,CMS:2012qbp} completed the Standard Model (SM). However, this does not rule out the existence of physics beyond the SM; instead, it merely restricts its possible scale and strongly encourages the experimental search for it.  New particles can be explored both directly through high-energy experiments and indirectly through low-energy precision frontier. Historically, indirect evidence for new particles has often come before their direct discoveries. Notably, the existence of the charm quark, the W boson, the top quark, and the Higgs boson was predicted based on indirect measurements such as Fermi interactions, kaon mixing, and electroweak precision observables. In this context, semileptonic $B$ meson decays are especially useful for conducting indirect searches for new physics (NP) due to the fact that they exhibit distinct and clear experimental signatures.

The flavor-changing neutral transitions, such as $b \rightarrow s\nu \bar\nu$ and $b \rightarrow s\ell\ell$, where $\ell$ represents a charged lepton, are highly suppressed in the SM  due to the Glashow-Iliopoulos-Maiani \cite{Glashow:1970gm} mechanism, making them a powerful tool for new physics searches beyond the SM  \cite{London:2021lfn,Capdevila:2023yhq}. In the SM the flavor-changing neutral current (FCNC) processes are forbidden at the tree level and take place only at loop level. Obviously these processes are inherently sensitive to any new physics. Rare semileptonic decays of mesons mediated by the FCNC are one of the promising areas to observe indirect NP effects. The rare $B$ meson decays with a neutrino pair in the final state, $B\rightarrow K^*\nu\bar{\nu}$ \cite{Altmannshofer:2009ma,Buras:2014fpa,Becirevic:2023aov,Felkl:2021uxi,Bause:2021cna,Browder:2021hbl,Becirevic:2024iyi} and $B^+\rightarrow K^+\nu\bar{\nu}$, are cleanest of them to search for new physics, due to their well-controlled theoretical uncertainty in the SM, and for this reason are rather sensitive to NP beyond SM. Many beyond-the-SM (BSM) scenarios predict a significant deviation of ${\rm Br}(B^+\rightarrow K^+\nu\bar{\nu})$ and ${\rm Br}(B\rightarrow K^*\nu\bar{\nu})$ with respect to their SM prediction and therefore those branching ratios can provide us with either a test of validity of a given model or with a constraint acceptable scenario of physics BSM. Experimentally, BABAR and Belle imposed upper limits on the $B\rightarrow K^*\nu\bar{\nu}$ decays and had a measurement of $B\rightarrow K\nu\bar{\nu}$ branching ratios, which are a few times higher than the SM predictions \cite{BaBar:2010oqg,Belle:2013tnz,BaBar:2013npw,Belle:2017oht}. Belle II was expected to achieve a measurement of the $B^+\rightarrow K^+\nu\bar{\nu}$, $B^+\rightarrow K^{*+}\nu\bar{\nu}$ and  $B\rightarrow K^{*}\nu\bar{\nu}$ branching ratios with a precision of about 10\%  \cite{Belle-II:2018jsg}.

Recently, using the dataset with an integrated luminosity of $362 fb^{-1}$, the Belle II experiment announced the first 3.5$\sigma$  evidence for the charged decay mode $B^+\rightarrow K^+\nu\bar{\nu}$, ${\rm Br}(B^+\rightarrow K^+\nu\bar{\nu})_{\rm Belle \, II}= (2.3\pm 0.7) \times 10^{-5}$  \cite{Belle-II:2023esi}, showing a 2.7$\sigma$ excess over the SM prediction ${\rm Br}(B^+\rightarrow K^+\nu\bar{\nu})_{\rm SM}= (4.29\pm 0.23) \times 10^{-6}$ \cite{Altmannshofer:2009ma,Buras:2014fpa,Becirevic:2023aov} where the tree-level contribution mediated by $\tau$ leptons has been subtracted \cite{Gabrielli:2024wys}. If the results are taken at face value, the following conclusions can be drawn: either lepton flavor universality is violated at the (multi)-TeV scale, or NP effects may be involved. This result is intriguing as it may suggest not only the existence of NP in the $b \rightarrow s\nu\bar{\nu}$ transitions but also the possible presence of new light states \cite{Altmannshofer:2023hkn}. The possibility of explaining the excess reported by Belle II in terms of NP has been considered in many different scenarios \cite{Allwicher:2023xba,Bause:2023mfe,Altmannshofer:2023hkn,McKeen:2023uzo,Fridell:2023ssf,Gabrielli:2024wys,Felkl:2023ayn,Wang:2023trd,He:2023bnk,Datta:2023iln,Ho:2024cwk,Chen:2024jlj,Hou:2024vyw,He:2024iju,Bolton:2024egx,Buras:2024ewl,Altmannshofer:2024kxb,Marzocca:2024hua,Hu:2024mgf,Allwicher:2024ncl,Berezhnoy:2023rxx,Calibbi:2025rpx,Lee:2025jky,Athron:2023hmz,Rosauro-Alcaraz:2024mvx}. Recently, research on these types of experimental signatures has intensified within the context of dark sectors.

The $B^+\rightarrow K^+\nu\bar{\nu}$ decay rate can be significantly influenced in models that introduce non-SM particles \cite{Belle-II:2023esi}, like leptoquarks \cite{Becirevic:2018afm}. Additionally, the $B$ meson might decay into a kaon and an invisible particle, such as an axion \cite{MartinCamalich:2020dfe} or a dark-sector mediator \cite{Filimonova:2019tuy}. There are other rare decays of the $B$ meson including the $\nu\bar{\nu}$ pair, and as mentioned above, one such channel would be $B\rightarrow K^{*}\nu\bar{\nu}$ decay, which involves the vector meson $K^*$ in the final state, while the kaon in the $B\rightarrow K\nu\bar{\nu}$ process is a scalar particle. The Belle search for this decay has shown no excess over the SM, so it is converted to an exclusion bound of $1.8\times 10^{-8}$ for ${\rm Br}(B\rightarrow K^{*}\nu\bar{\nu})$ \cite{Belle:2017oht}. 

In the last few years, the existence of dark sectors has emerged as a compelling theoretical framework. The simplest dark sector may be referred to as the dark photon model \cite{Holdom:1985ag,Fabbrichesi:2020wbt}, where a new Abelian gauge boson is added to the particle content of the SM. Dark sectors generally contain one or more mediator particles that interact with the SM through a portal. The portal relevant for interactions between the dark sector and the SM depends on the mediator's spin and parity. The portal can be a scalar, a pseudoscalar, a fermion, or a vector. The gauge and Lorentz symmetries of the SM significantly restrict the possible ways the mediator can couple to the SM. In particular, light dark sectors with masses ranging from few MeV to few GeV have triggered a lot of attention, and the Belle II experiment is known to be highly sensitive to many such scenarios.

The dark photon scenario \cite{Fayet:1980rr,Fayet:1990wx,Holdom:1985ag} has been extensively analyzed in the literature, mainly in its massive case, and is the focus of many ongoing experimental searches \cite{Fabbrichesi:2020wbt}. In the context of astrophysics and cosmology, massless dark photons have been studied as  mechanisms for long-range interactions among dark matter components \cite{Gradwohl:1992ue,Ackerman:2008kmp,Fan:2013tia,Foot:2014uba,Acuna:2020ccz}. Additionally, massless dark photons have been employed to understand the Standard Model's flavor hierarchy problem, flavor-changing neutral currents, and phenomena in kaon physics \cite{Gabrielli:2013jka,Gabrielli:2016vbb,CMS:2020krr,Gabrielli:2016cut,Fabbrichesi:2017vma}.

Dark photon models can be classified into two separate types depending on whether the associated field quanta are massive or massless. These possibilities lead to distinctly different experimental signatures: a massless dark boson does not have tree-level couplings to any SM fermions and thus can only interact with ordinary matter through operators of dimension higher than 4 \cite{Holdom:1985ag,Fabbrichesi:2020wbt,Dobrescu:2004wz}. In contrast, massive dark bosons can interact with ordinary matter through a renormalizable dimension-4 operator involving an arbitrarily small charge. These two categories are distinct since the  massless case cannot be achieved as a limit of massive dark photon models.  

In the present work, we study the compatibility of the recent Belle II result for ${\rm Br}(B^+\rightarrow K^+\nu\bar{\nu})$ with new physics in the form of a two-Higgs-doublet model extended with $U(1)_D$. The part of the parameter space that satisfies the ${\rm Br}(B^+\rightarrow K^+\nu\bar{\nu})$ constraint will be further checked against the  ${\rm Br}(B^0\rightarrow K^{*0}\nu\bar{\nu})$ bound, which would reduce the allowed space significantly. Our paper is organized as follows. After the two-Higgs-doublet model extended with a dark gauge group is briefly outlined in Sec. \ref{model}, we present the decay widths of $B\to K^{(*)}\nu\bar{\nu}$ processes within the framework of $U(1)_D$ extended two-Higgs-doublet models in Sec. \ref{nes}. The next section is devoted to the numerical analysis of branching ratios mentioned above. The last section contains our discussions and conclusions.

\section{$U(1)_D$ extended Two-Higgs Doublet Models}
\label{model}

In this section we outline the basic elements of the theoretical framework and we refer to Refs. \cite{Elpe:2022hqp,Campos:2017dgc} for further details. Then a numerical analysis of the process $B\rightarrow K^{(*)}\nu\bar{\nu}$ will follow.

Although its definition is rather broad, dark sector physics can be effectively and systematically investigated by using specific portal interactions as a guide. Dark sectors typically contain one or more mediator particles that connect to the SM through a portal. The interactions between the SM and  dark sector could be controlled through various portals, classified further based on types of particles involved. A vector portal, requiring a dark vector boson coupling with the SM particles through dimension-4 operators, is one of such commonly studied scenarios.

The gauge group of the Abelian vector portal is $U(1)_D$; the mediator originating from this additional gauge group is a vector boson, known as the dark $Z$. Even though it is taken to be massive, the vector boson (dark photon/dark $Z$) resembles the SM photon because it is based on a U(1) symmetry\footnote{As far as conventions about the naming of this new vector boson, as dark photon or dark $Z$, are concerned, there might be different usages in the literature. However, the guiding principle would be that if the new vector boson has {\it axial} vector couplings it resembles the SM $Z$ and thus would be called dark $Z$, rather than dark photon. We will follow this convention and call it dark $Z$ in the rest of the paper.}; however, unlike the photon, it mediates a new kind of fundamental force, known as the  dark force, which might have some detectable effects at colliders as well as low energy with high-luminosity experiments. The vector portal fundamentally involves an interaction through kinetic mixing between the SM $U(1)_Y$ hypercharge field $Y_\mu$ and the $U(1)_D$ dark sector field $X_\mu$. This kinetic mixing interaction ($\frac12 \sin\!\epsilon\ Y_{\mu\nu}X^{\mu\nu}$) is invariant under both the $U(1)_Y$ and $U(1)_D$ gauge groups. The strength of the coupling is defined by the kinetic mixing parameter $\sin\epsilon$, which is a free parameter. Since it behaves as a small perturbation to the SM, to be consistent with observations, it is generally expected to have a small value. The phenomenology of the vector portal represents a broad class of well-motivated models, including scenarios where the mediator preferentially couples to baryonic, leptonic, or baryon number minus lepton number ($B-L$) conserving currents.

In its original form, there exists only a nonzero mixing between the photon field and the dark vector field, which results in limited access to the SM fermions without involving any additional gauge symmetry. This picture can be extended if the dark vector field is a gauge field of a local $U(1)$ symmetry under which the SM fields are charged. Indeed, baryon number minus lepton number ($B-L$) is assigned to the SM fields plus right-handed neutrinos as the new charges, and one can show that the theoretical framework with $G_{\rm SM}\times U(1)_D$ gauge symmetry would be free of gauge anomalies.

A popular way to extend the SM is through its scalar sector while keeping the dark sector minimal (specifically, assuming an Abelian dark gauge group $U(1)_D$). This is achieved by introducing an additional $SU(2)$ doublet, leading to the so-called two-Higgs-doublet models (2HDMs) \cite{Lee:1973iz}. In its general form, the 2HDM struggles with the FCNC problem. To resolve this issue, a discrete symmetry is typically required. Furthermore, neutrino masses, which are among the most important observational evidence for new physics, are typically not addressed in the 2HDM. In this study, a gauge solution will be employed to address the FCNC problem. The idea here is to obtain anomaly-free, conventional 2HDM scenarios without FCNC by adding an Abelian $U(1)_D$ gauge group to the 2HDM ($G_{\rm 2HDM}\bigotimes U(1)_D$). The particle content of the model will include right-handed neutrinos, which will not only address the neutrino mass issue but also make it possible to satisfy the new anomaly equations that arise in the 2HDM under the $U(1)_D$ group. The details of the scenarios have been extensively examined in previous studies \cite{Lee:2013fda,Campos:2017dgc,Lindner:2018kjo}. We focus solely on the relevant part of the model which we need in the present study. The Lagrangian of the considered model is 
	\begin{eqnarray}
		{\cal L} &=&-\frac14 W_{3\mu\nu} W_3^{\mu\nu}-\frac14 Y_{\mu\nu} Y^{\mu\nu}-\frac14 X^0_{\mu\nu}
		X^{0\mu\nu}
		-\frac12 \sin\!\epsilon\ X^0_{\mu\nu}Y^{\mu\nu} +  \frac12 m_{X}^2 X^0_\mu X^{0\mu}\nonumber \\
		&&+ (D_\mu \phi_1)^\dagger(D^\mu \phi_1) + (D_\mu \phi_2)^\dagger(D^\mu \phi_2)  + (D_\mu \phi_s)^\dagger(D^\mu \phi_s) +\sum_i \bar{f}_i i{\slashchar D} f_i\,.
		\label{Ltotal}
\end{eqnarray}
Here, $X^0_\mu$ represents the gauge field of $U(1)_D$ while $Y_\mu$ and $W_{3\mu}$ are the usual weak hypercharge field of $U(1)_Y$ and the third of weak gauge fields, respectively. $\phi_1$ and $\phi_2$  are the usual scalar doublets under $SU(2)_L$ with $\langle \phi_i\rangle=v_i/\sqrt{2}$, while $\phi_s$ is an additional scalar singlet with $\langle \phi_s\rangle=v_s/\sqrt{2}$ to explain the neutrino masses.  $\sin\epsilon$  indicates the strength of the kinetic mixing  between $U(1)_Y$ and $U(1)_D$. $m_X$ is the Stueckelberg mass parameter \cite{Stueckelberg:1938hvi,Elpe:2022hqp} for the dark gauge field $X^0_\mu$. $D_\mu =\partial_\mu-ig_W\, t^3\, W_{3\mu} - i g_Y Y Y_\mu -i g_D  Q^{'} X^0_\mu + \dots$ is the covariant derivative with $t^3$, $Y$, and $Q^{'}$  being the $SU(2)_L$ generator, weak hypercharge, and dark $U(1)_D$ charges of the fields, respectively.  Here, $g_W$, $g_Y$, and $g_D$ refer to the corresponding gauge coupling constants.
In the literature, the Stueckelberg extension of the SM with and without kinetic mixing has been thoroughly analyzed \cite{Feldman:2007wj,Kors:2004dx}. The two-Higgs doublet model, augmented by an additional $U(1)$ gauge symmetry, has been examined in Ref. \cite{Panotopoulos:2011xb} by considering the Stueckelberg contribution to the mass terms.

\begin{table*}[h]
	\sisetup{table-align-text-pre=false, table-align-text-post=false} 
	\caption{Dark quantum charges of the fields under $U(1)_D$, adapted from Ref. \cite{Campos:2017dgc}.} \label{ChargeTable}
	\addtolength{\tabcolsep}{0.7pt}
	\begin{tabularx}{\textwidth}{@{\extracolsep{\fill}} lrrrrrrrr}
		\hline \hline \; Fields & $u_R$ & $d_R$ & $Q_L$ & $L_L$ & $e_R$ & $\nu_R$ &		$\phi_2$ & $\phi_1$     \\\hline \\[-3.1ex]
		$\;$  Dark charges & $Q_u'$ & $Q_d'$ & $\frac{Q_u'+Q_d'}{2}$ & $-\frac{3(Q_u'+Q_d')}{2}$ & $-(2Q_u'+Q_d')$ & $-(Q_u'+2Q_d')$ & $\frac{Q_u'-Q_d'}{2}$ & $\frac{5Q_u'+7Q_d'}{2}$ \\[0.2em]\hline
		$\;$  Model A&$\frac{1}{2}$ &$-\frac{1}{2}$ &$0$&$0$&$-\frac{1}{2}$&$\frac{1}{2}$&$\frac{1}{2}$&$-\frac{1}{2}$ \\
		$\;$  Model B&\hphantom{-} $\frac{1}{2}$ &$\frac{1}{2}$ &$0$&$0$&$\frac{1}{2}$&$-\frac{1}{2}$&$-\frac{1}{2}$&$\frac{1}{2}$ \\
		$\;$  Model C&$\frac{1}{4}$ &$-\frac{1}{2}$ &$-\frac{1}{8}$&$\frac{3}{8}$&0&$\frac{3}{4}$&$\frac{3}{8}$&$-\frac{9}{8}$ \\
		$\;$ Model D &$\frac{1}{2}$&0&$\frac{1}{4}$&$-\frac{3}{4}$&$-1$&$-\frac{1}{2}$& $\frac{1}{4}$&$\frac{5}{4}$    \\
		$\;$  Model E&0&$\frac{1}{2}$&$\frac{1}{4}$&$-\frac{3}{4}$&$-\frac{1}{2}$&$-1$&$-\frac{1}{4}$&$\frac{7}{4}$ \\
		$\;$  Model F &$\frac{2}{3}$&$\frac{1}{3}$&$\frac{1}{2}$&$-\frac{3}{2}$&$-\frac53$&$-\frac{4}{3}$&$\frac{1}{6}$&$\frac{17}{6}$ \\
		$\;$ Model G &$-\frac{1}{6}$&$\frac{1}{3}$&$\frac{1}{12}$&$-\frac{1}{4}$&0&$-\frac{1}{2}$&$-\frac{1}{4}$&$\frac{3}{4}$    \\
		$\;$  Model $B-L$&$\frac{1}{6}$&$\frac{1}{6}$&$\frac{1}{6}$&$-\frac{1}{2}$&$-\frac{1}{2}$& $-\frac{1}{2}$&0&1\\
		\hline\hline
	\end{tabularx}
\end{table*}

The comprehensive discussion of the scalar sector and the anomaly cancellation mechanisms of the model can be found in Ref. \cite{Campos:2017dgc}. Certain relevant parts will be reiterated here. For instance, the dark charge assignments of the SM fields that satisfy the anomaly conditions are provided in Table~\ref{ChargeTable}. Since the original gauge basis in the neutral sector, i.e., $(Y_\mu,W_{3\mu},X^0_\mu)$, is not diagonal, the physical basis, denoted as  $(A_\mu, Z_\mu, Z^{'} _\mu)$, can be derived through three successive rotations (the first one to put kinetic terms into the canonical form and then the usual Weinberg angle rotations with $\theta_W$ and lastly a rotation by an angle $\xi$ to eliminate the mass mixing terms) (like given in Ref. \cite{Elpe:2022hqp} but with additional contribution from the scalar singlet). After performing the rotations, one can finally obtain eigenvalues corresponding to the mass squares of the photon ($A_\mu$), the dark $Z$ ($Z^{'} _\mu$), and the electroweak neutral boson ($Z_\mu$), respectively:
 \begin{eqnarray}
	M_A^2 &=& 0, \\
	m_{Z^'}^2 =&& \sec^2\epsilon\, \cos^2\xi \, m_X^2 + \sec^2\epsilon \cos^2\xi \left(v^2 \left(\cos^2\beta \, \widetilde{Q}_{\phi_1}^2 + \sin^2\beta \, \widetilde{Q}_{\phi_2}^2\right) + v_s^2 \widetilde{Q}_{\phi_s}^2\right) \nonumber \\ 
	&&+ 2 v \sec\epsilon \cos\xi \, m_{Z_0} \left(\cos^2\beta \, \widetilde{Q}_{\phi_1} + \sin^2\beta \, \widetilde{Q}_{\phi_2}\right) \left(\sin\xi - \cos\xi \sin\theta_W \tan\epsilon\right)\nonumber\\
	&& + m_{Z_0}^2 \left(\sin\xi - \cos\xi \sin\theta_W \tan\epsilon\right)^2, \label{eqn:MAp}\\
m_{Z}^2 =&& \sec^2\epsilon\, \sin^2\xi \, m_X^2 + \sec^2\epsilon \sin^2\xi \left(v^2 \left(\cos^2\beta \, \widetilde{Q}_{\phi_1}^2 + \sin^2\beta \, \widetilde{Q}_{\phi_2}^2\right) + v_s^2 \widetilde{Q}_{\phi_s}^2\right) \nonumber \\ 
&&- 2 v \sec\epsilon \sin\xi \, m_{Z_0} \left(\cos^2\beta \, \widetilde{Q}_{\phi_1} + \sin^2\beta \, \widetilde{Q}_{\phi_2}\right) \left(\cos\xi + \sin\xi \sin\theta_W \tan\epsilon\right)\nonumber\\
&& + m_{Z_0}^2 \left(\cos\xi + \sin\xi \sin\theta_W \tan\epsilon\right)^2.\label{eqn:MZ}
\end{eqnarray}  

Here, $\tan\beta\equiv v_2/v_1$ (with $\sqrt{v_1^2+v_2^2} = v = 246\, \textrm{GeV}$). $Q_{\phi_i}^{'}$ is the dark charge of the scalar doublet $\phi_i\,,\ i=1,2$. $m_{Z_0}=\frac12 \sqrt{g_Y^2+g_W^2}\,v$ is the mass of the SM $Z$ boson. The parameter $\sin\epsilon$  describes the strength of the kinetic mixing  between $U(1)_Y$ and $U(1)_D$. From now on, we define $g_D Q^{'}_{\phi_i} \equiv \widetilde{Q}_{\phi_i}^{'}$ for later convenience\footnote{Since the gauge coupling $g_D$ 
	and the free charges of the model ($Q_u^{'}$ and $Q_d^{'}$) 
	enter the vertex factors and other relevant terms via the covariant derivative as a product, 
	the tilde parameters have been defined accordingly.} and  $\widetilde{Q}_{\phi_s}^{'}= \widetilde{Q}_{\phi_1}^{'}-\widetilde{Q}_{\phi_2}^{'}$.

The rotation angle $\xi$ should satisfy $\tan 2\xi = 2b/(a-c)$ where the parameters $a,b,$ and $c$ are given (like given in Ref. \cite{Elpe:2022hqp} with nonzero $v_s$),
\begin{eqnarray}
	\label{eq:xieqn}
	a &=& m_{Z_0}^2, \nonumber\\
	b &=& m_{Z_0}^2\left(\cos^2\!\beta\ \Delta_1 + \sin^2\!\beta\ \Delta_2\right)\,,\\
	c &= &  \sec^2\!\epsilon (m_X^2+v_s^2 \widetilde{Q}_{\phi_s}^2) + m_{Z_0}^2 \left(\cos^2\!\beta\ \Delta_1^2 + \sin^2\!\beta\ \Delta_2^2\right)\,. \nonumber
\end{eqnarray}
Here, the functions $\Delta_i$ are defined as follows:
\begin{eqnarray}
	\label{eq:Deltai}
	\Delta_i = \sin\!\theta_W \tan\!\epsilon - \frac{v\,\sec\!\epsilon}{m_{Z_0}}\, \widetilde{Q}_{\phi_i}^{'}\,,\qquad i=1,2.
\end{eqnarray}
A compelling feature of this model is its ability to explain the small yet nonzero Dirac masses observed in neutrinos through the seesaw mechanism. In the neutrino sector, we can incorporate the Yukawa Lagrangian terms $-y_D\overline{L_L}\widetilde{\phi_2}\nu_R-y_M\overline{\nu_R^c}\widetilde{\phi_s}\nu_R$, which generate both Dirac and Majorana mass terms. These terms are governed by the vacuum expectation values $v_2$ and $v_s$, respectively, in conjunction with their corresponding Yukawa couplings $y_D$ and $y_M$. 

Once the diagonalization of the mass matrix is performed, the effective neutrino mass emerges in the canonical seesaw form $-m_D^Tm_R^{-1}m_D$, where $m_D=y_D v_2/(2\sqrt{2})$ represents the Dirac mass term and $m_R=y_M v_s/(2\sqrt{2})$ corresponds to the Majorana mass term. For our numerical analysis, we fix the vacuum expectation value of the scalar singlet $v_s$ at the 1~TeV scale. Under these conditions, with $y_M \simeq 1$, the right-handed neutrinos acquire masses of approximately 350~GeV. Simultaneously, the left-handed neutrinos $\nu_L$ attain masses of 0.1~eV when $y_D = 2.4 \times 10^{-6}$ and $\tan\beta = 2$, consistent with experimental constraints on neutrino masses\cite{Planck:2018vyg}.

\begin{table*}[htb]
	\caption{The relevant fermion-fermion-$Z/Z^{'}$ vertex factors for $B\to K^{(*)}\nu\bar{\nu}$ decays in the Two-Higgs Doublet Models extended with a dark $U(1)_D$ group. A shorthand notation is used for the trigonometric expressions. For example, $(s_\xi\,, t_\epsilon)$ stand for $(\sin \xi\,, \tan\!\epsilon$) and similar for the others.}
	\label{tab:vertex}
	\begin{tabularx}{0.99\linewidth}{X c >{\centering\arraybackslash}p{1.2cm}>{\centering\arraybackslash}c >{\centering\arraybackslash}c }
		\hline\hline
		Vertices &   $C_V^{f}$ & &  $C_A^{f}$  \\
		\hline\\[-1.5em]
		\includegraphics[scale=0.8]{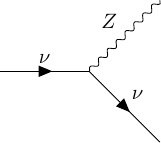} 
		&
		\makecell[l]{\vspace*{1.5cm}
			{ $-\frac{ s_\xi  \br{7\widetilde{Q}^{'}_d + 5 \widetilde{Q}^{'}_u} }{4c_\epsilon} 
				+\frac{e \br{c_\xi + s_\xi t_\epsilon s_W}}{4 c_W s_W}$}}
		& &
		\makecell[l]{\vspace*{1.5cm}
			{ $ -\frac{ s_\xi  \br{ \widetilde{Q}^{'}_d -  \widetilde{Q}^{'}_u} }{4c_\epsilon}
				-\frac{ e \br{c_\xi + s_\xi t_\epsilon s_W} }{4 c_W s_W}$}}\\[-10mm]
		\includegraphics[scale=0.8]{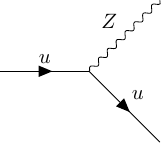} 
		&
		\makecell[l]{\vspace*{1.5cm}
			{$ \frac{  s_\xi  \br{ \widetilde{Q}^{'}_d + 3 \widetilde{Q}^{'}_u} }{4 c_\epsilon}
				-\frac{ e c_\xi  \br{8s_W^2-3}+ 5e s_\xi t_\epsilon s_W}{12 c_W s_W}$
		}}
		& &
		\makecell[l]{\vspace*{1.5cm}
			{$ -\frac{ s_\xi  \br{ \widetilde{Q}^{'}_d -  \widetilde{Q}^{'}_u}}{4c_\epsilon}
				-\frac{ e  \br{c_\xi + s_\xi t_\epsilon s_W}}{4 c_W s_W}$
		}}\\[-10mm]
		\includegraphics[scale=0.8]{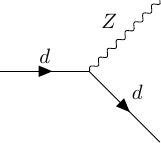} 
		&
		\makecell[l]{\vspace*{1.5cm}
			{$ \frac{ s_\xi  \br{ 3 \widetilde{Q}^{'}_d + \widetilde{Q}^{'}_u} }{4c_\epsilon}
				+\frac{e c_\xi  \br{4s_W^2-3} +  e s_\xi t_\epsilon s_W}{12 c_W s_W}$}}
		& &
		\makecell[l]{\vspace*{1.5cm}
			{$ \frac{ s_\xi \br{ \widetilde{Q}^{'}_d -  \widetilde{Q}^{'}_u} }{4c_\epsilon } 
				+\frac{  e  \br{c_\xi + s_\xi t_\epsilon s_W}}{4 c_W s_W} $
		}} \\[-8mm]
		\hline
		&   $C_V^{'f}$ & &  $C_A^{'f}$ \\
		\hline\\[-6mm]
		\includegraphics[scale=0.8]{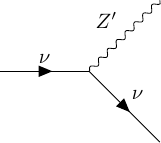} 
		&
		\makecell[l]{\vspace*{1.5cm}{$ -\frac{ c_\xi \br{7 \widetilde{Q}^{'}_d + 5 \widetilde{Q}^{'}_u} }{4c_\epsilon } 
				-\frac{  e  \br{s_\xi - c_\xi t_\epsilon s_W} }{4 c_W s_W} $
		}}
		& &
		\makecell[l]{\vspace*{1.5cm}{$ -\frac{ c_\xi \br{\widetilde{Q}^{'}_d - \widetilde{Q}^{'}_u}}{4c_\epsilon } 
				+\frac{  e  \br{s_\xi - c_\xi t_\epsilon s_W} }{4 c_W s_W} $
		}}\\[-10mm]
		\includegraphics[scale=0.8]{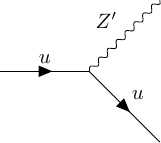} 
		&
		\makecell[l]{\vspace*{1.5cm}{$ \frac{  c_\xi  \br{ \widetilde{Q}^{'}_d + 3 \widetilde{Q}^{'}_u}}{4c_\epsilon }
				+ \frac{ e  s_\xi  \br{8s_W^2-3}- 5e c_\xi t_\epsilon s_W}{12 c_W s_W}$
		}}
		& &
		\makecell[l]{\vspace*{1.5cm}{$ -\frac{ c_\xi  \br{ \widetilde{Q}^{'}_d -  \widetilde{Q}^{'}_u} }{4c_\epsilon }
				+\frac{ e  \br{s_\xi - c_\xi t_\epsilon s_W}}{4 c_W s_W}$
		}}\\[-10mm]
		\includegraphics[scale=0.8]{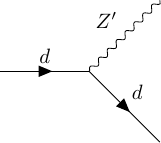} 
		&
		\makecell[l]{\vspace*{1.5cm}{$ \frac{  c_\xi \br{ 3 \widetilde{Q}^{'}_d + \widetilde{Q}^{'}_u}}{4 c_\epsilon } 
				-\frac{ e  s_\xi  \br{4s_W^2-3} - e c_\xi t_\epsilon s_W}{12 c_W s_W} $
		}}
		& &
		\makecell[l]{\vspace*{1.5cm}{$   \frac{ c_\xi  \br{ \widetilde{Q}^{'}_d -  \widetilde{Q}^{'}_u}}{4c_\epsilon }
				- \frac{ e  \br{s_\xi - c_\xi t_\epsilon s_W}}{4 c_W s_W}$
		}}\\[-8mm]
		\hline\hline
	\end{tabularx}	
	\end{table*}

\section{$B\to K^{(*)}\nu\bar{\nu}$ decay in $U(1)_D$-extended 2HDM}
\label{nes}

After describing the necessary ingredients of the theoretical framework in the previous section, we now delve into the details of the matrix element for the \( b \rightarrow s\nu\bar{\nu} \) transition within the \( U(1)_D \)-extended two-Higgs-doublet model. For convenience, we give a list of the vertex factors for fermion-fermion-$Z/Z^{'}$ interactions that are relevant to the $B\rightarrow K^{(*)}\nu\bar{\nu}$ decays in Table~\ref{tab:vertex}, according to the parametrization given by the following Lagrangian terms:
\begin{align}
	\mathcal{L} \supset \bar{f}(C_V^f \gamma^\mu + C_A^f \gamma^\mu \gamma^5)f Z_\mu + \bar{f} (C_V^{'f} \gamma^\mu + C_A^{'f} \gamma^\mu \gamma^5) f Z^{'} _\mu\,.
\end{align}

The vertex factors of the models, namely, models A, B, C, D, E, F, G, and $B-L$ (listed in Table~\ref{ChargeTable}), can be read off from the entries in Table~\ref{tab:vertex} by plugging the values of the corresponding charges $Q'_u$ and  $Q'_d$, given in Table~\ref{ChargeTable}. By taking the appropriate limiting values, $Q'_{u,d}\to 0$ (or $g_D\to 0$) and  $\epsilon\to 0$\ (which makes $\xi\to 0$, seen from Eqs.\,(\ref{eq:xieqn}) and (\ref{eq:Deltai})), the vertex factors in Table~\ref{tab:vertex} can be reduced to  their SM forms. In this limit, it is clear that all dark $Z$ vertices vanish, as expected.

The $b\to s$ transition happens at one-loop level, receiving contributions from the so-called self-energy diagrams, triangle diagrams involving \( Z \) and dark $Z$ exchanges, and box diagrams. At this point, two key observations are worth noting: a) the model introduces right-handed neutrino currents, and b) neither $Z$ nor the dark $Z$ contributes to the box diagrams. The contribution from diagrams involving \( Z \)-boson exchange can be expressed as
\begin{align}
	{\cal M}_{Z}= \frac{V_{tb} V_{ts}^* G_F \, \alpha _{em} \, m_W^2}{2\sqrt{2} \pi m_{Z}^2  }  \left(C_{LL}^{Z} {\cal O}_{LL} + C_{LR}^{Z} {\cal O}_{LR}\right), \label{eqn:Zbosonexc}
\end{align}
while the matrix element for dark $Z$ exchange takes the form
\begin{align}
	{\cal M}_{Z^{'} }= \frac{V_{tb} V_{ts}^* G_F \, \alpha _{em} \, m_W^2}{2\sqrt{2} \pi (m_{Z^{'}}^2 - q^2) }  \left(C_{LL}^{Z^{'} } {\cal O}_{LL} + C_{LR}^{Z^{'} } {\cal O}_{LR}\right). \label{eqn:darkphtnexc}
\end{align}
Here, the coefficients are defined as
\begin{align}
	C_{LL(R)}^{Z} = &\ C_{LL(R)}^{TZ}+C_{LL(R)}^{SZ},  \\
	C_{LL(R)}^{Z^{'} } = &\ C_{LL(R)}^{TZ^{'} }+C_{LL(R)}^{SZ^{'} }, 
\end{align}
where the superscripts \( T \) and \( S \) denote contributions from triangle and self-energy diagrams, respectively. Lastly, the matrix element for the box diagrams can be expressed as
\begin{align}
	\label{eqn:mbox}
	{\cal M}^{\text{Box}}= \frac{V_{tb} V_{ts}^* \, G_F \, \alpha _{em} \, m_W^2}{2\sqrt{2} \pi}\   C_{LL}^{Box} {\cal O}_{LL} 
\end{align}
Explicit expressions for these $C$ coefficients are provided in Appendix \ref{sec:appenA}. The operators \( {\cal O}_{LL} \) and \( {\cal O}_{LR} \) correspond to the local ones
\begin{eqnarray}
 {\cal O}_{LL}  &=& \bar{s}\gamma_{\mu}(1-\gamma_5)b \, \bar{\nu}\gamma_{\mu}(1-\gamma_5)\nu\,,   \nonumber\\ 
 {\cal O}_{LR}  &=&
 \bar{s}\gamma_{\mu}(1-\gamma_5)b \, \bar{\nu}\gamma_{\mu}(1+\gamma_5)\nu\,. 	\nonumber
\end{eqnarray}
Using the contributions at the quark level given in Eqs.~(\ref{eqn:Zbosonexc}), (\ref{eqn:darkphtnexc}), and (\ref{eqn:mbox}), the hadronic matrix element for the \( B\rightarrow K^{(*)}\nu\bar{\nu} \) decays can be calculated by sandwiching the quark current between the initial \( B \) meson and the final \( K^{(*)} \) meson states. The hadronic matrix elements appearing in $B\to K^{(*)}\nu\bar{\nu}$ decays are parametrized in terms of form factors, which are further classified as scalar, psudoscalar, vector, and axial vector.

The matrix elements for the $B\to K^{(*)}$ transitions are
\begin{eqnarray}
	\braket{K\left(p_K \right)|\bar{s}\gamma_\mu b|B\left(p_B \right)} &=&\left[ (p_B + p_K)_\mu - \frac{m_B^2 - m_K^2}{q^2} q_\mu \right]f_+(q^2) + \frac{m_B^2 - m_K^2}{q^2} q_\mu f_0(q^2),\nonumber\\	
	\braket{K^*\left(p_{K^*} , \varepsilon \right)|\bar{s}\gamma_\mu b|B\left(p_B \right)} 
	&=& \epsilon _{\mu \nu \rho \sigma} \varepsilon^{* \nu} p_B^\rho p_{K^*}^\sigma \frac{2V(q^2)}{m_B+ m_{K^*}},\nonumber \\
	\braket{K^*\left(p_{K^*} , \varepsilon \right)|\bar{s}\gamma_\mu \gamma_5  b|B\left(p_B \right)} &=& \mi \varepsilon^{*\nu} \Bigg[ \eta_{\mu \nu} (m_B+ m_{K^*})A_1(q^2) - \frac{(p_B + p_{K^*})_\mu  q_\nu}{m_B+ m_{K^*}} A_2(q^2)\nonumber \\
	&& \left.
	- q_\mu q_\nu \frac{2 m_{K^*}}{q^2} (A_3(q^2)-A_0(q^2)) \Bigg]\right.\nonumber\,,
	\end{eqnarray}
where the scalar and vector form factors for the \( B\to K \) transition are \( f_0(q^2) \) and \( f_+(q^2) \), while the vector and axial form factors for \( B\to K^* \) are \( A_0(q^2),  A_1(q^2),  A_2(q^2), A_3(q^2),\) and \(V(q^2)\). Among these, \( f_0, A_0\), and \(A_3\) do not contribute to  \( B\rightarrow K^{(*)}\nu\bar{\nu} \) decays due to having negligible neutrino masses. The remaining set of form factors can be chosen as $(f_+, A_1,A_{12},V)$ after employing the following relation:
\begin{eqnarray}
	A_2 = -\frac{(m_B+m_{K^*}) \left[A_1 (m_B+m_{K^*})
		\left(-m_B^2+m_{K^*}^2+q^2\right)+16 A_{12} m_B
		m_{K^*}^2\right]}{m_B^4-2 m_B^2
		\left(m_{K^*}^2+q^2\right)+\left(m_{K^*}^2-q^2\right)^2}
\end{eqnarray}

The differential decay widths for the exclusive decays \( B\rightarrow K\nu\bar{\nu} \) and \( B\rightarrow K^*\nu\bar{\nu} \) can be cast into the following compact forms:
\begin{eqnarray}
	\frac{d\Gamma (B\rightarrow K\nu\bar{\nu})}{dq^2} 
	&=& \frac{1}{512\pi^3 m_B^3} f_+^2(q^2) (C_{LL}^2 + C_{LR}^2) \lambda ^{3/2},\\  
	\frac{d\Gamma (B\rightarrow K^*\nu\bar{\nu})}{dq^2} 
	&=& \frac{\lambda^{1/2}
		\left(C_{LL}^2+C_{LR}^2\right)}
	{256 \pi ^3 m_B^3
		\left(m_B+m_{K^*}\right)^2}
	\left[\lambda\, q^2
	V^2+A_1^2 q^2 \left(m_B+m_{K^*}\right)^4
	\right. \nonumber \\
	&& \left. +32
	A_{12}^2 m_B^2 m_{K^*}^2
	\left(m_B+m_{K^*}\right)^2\right],
\end{eqnarray}
where \( \lambda(m_B^2,m_{K^{(*)}}^2,q^2)=m_B^4+m_{K^{(*)}}^4+q^4-2m_B^2m_{K^{(*)}}^2-2m_B^2q^2-2m_{K^{(*)}}^2q^2 \), and
\begin{align}
	C_{LL(R)}= \frac{V_{tb} V_{ts}^* G_F \, \alpha _{em} \, m_W^2}{2\sqrt{2} \pi}  \left(\frac{C_{LL(R)}^{Z}}{m_{Z}^2 }  + \frac{C_{LL(R)}^{Z^\prime}}{m_{Z^{'}}^2-q^2 }   \right).
\end{align}
Here, $q^2=(p_B-p_{K^{(*)}})^2$ and the explicit forms of the coefficients $C_{LL(R)}^{Z}$ and $C_{LL(R)}^{Z^\prime }$ are given in Appendix \ref{sec:appenA}. In the numerical study, we demonstrate that treating the resonant production of the dark $Z$ properly plays very important role to enhance the new physics effects in ${\rm Br}(B\to K^{(*)}\nu\bar{\nu})$. The width affects can placed with the use of the Breit–Wigner formula
\begin{align}
	\frac{1}{q^2-m_{Z^{'}}^2} \rightarrow \frac{1}{q^2-m_{Z^{'}}^2+i\Gamma_{Z^{'} }m_{Z^{'}}},  
\end{align}
where \( \Gamma_{Z^{'}} \) denotes the decay width of the dark $Z$ into the relevant fermion pairs. At this point, it would be useful to provide an analytical formula for the decay width of $Z^{'}$ when $m_{Z^{'}}>2 m_f\ (f=\nu_\ell,e,\mu,\tau,u,d,c,s,b)$
\begin{eqnarray}
	\Gamma_{Z^{'}}(Z^{'}
\to f\bar{f}) = \sum_{f}\frac{N_f m_{Z^{'}}}{48\pi}v_f
\left\{v_f^2\lvert C_V^{'f}\rvert^2+\frac{3-v_f^2}{2}\lvert C_A^{'f}\rvert^2 \right\}\end{eqnarray}
where $v_f=\sqrt{1-4m_f^2/m_{Z^{'}}^2}$ and $N_f$ is equal to 3 for quarks and 1 for the others. The fermion-fermion-$Z^{'}$ couplings, $C_V^{'f}$ and $C_A^{'f}$, are given explicitly in Table~\ref{tab:vertex}. 

In the numerical analysis of both decays, the following \(z\) parametrizations for the  form factors $(f_+, F_i= A_1,A_{12},V)$  are employed \cite{Becirevic:2023aov, Bharucha:2015bzk}:
\begin{eqnarray}
	f_+\left(q^2\right) &=& P_+\left(q^2\right) \sum_{k=0}^{N-1}a_k^+\left[ z(q^2)^k-(-1)^{k-N}\frac{k}{N}z(q^2)^N\right]\,,\nonumber\\
F_i\left(q^2\right) &=& P_i\left(q^2\right) \sum_k a_k^i\left[z\left(q^2\right)-z(0)\right]^k
\end{eqnarray}
Here, $P_{i,+}\left(q^2\right) = \left(1-q^2/M_{i,+}^2\right)^{-1}$ with $M_+= 5.4154$ GeV, and the other mass parameters are 
\[ M_i= \begin{cases} 
	5.415\, {\rm GeV} & F_i=V \\
	5.829\, {\rm GeV} & F_i=A_1,A_{12}\,.\\ 
\end{cases}
\]
The $z$ function is defined as
\begin{align}
	z(q^2) =\frac{\sqrt{t_+-q^2}-\sqrt{t_+-t_0}}{\sqrt{t_+-q^2}+\sqrt{t_+-t_0}}
\end{align}
where $t_\pm = (m_B\pm m_{K^{(*)}})^2$ and $t_0=t_+\left[1-\left(1-t_-/t_+\right)^{\frac{1}{2}} \right]$. Lastly, the fit parameters $a_k^{+,i}$ are used from Refs. \cite{Becirevic:2023aov} and \cite{Bharucha:2015bzk}.

\section{Numerical Analysis of $B\to K^{(*)}\nu\bar{\nu}$}
\label{num}
In light of the analytical calculations presented in the previous section, a numerical analysis is required to examine the selected models at the given sample points. More importantly, we need to explore the parameter space to determine whether there exist scenarios that simultaneously accommodate the ${\rm Br}(B\to K\nu\bar{\nu})$ \cite{Belle-II:2023esi} and ${\rm Br}(B\to K^*\nu\bar{\nu})$ \cite{Belle:2017oht} data from the Belle Collaboration.

The model primarily includes the following five parameters: 
$\widetilde{Q}_u^{'} $, 
$\widetilde{Q}_d^{'}$, 
$\sin\epsilon$, $\tan\beta$, and $M_{Z^{'}}$.  The $\widetilde{Q}$ parameters can, in principle, be large, provided that they do not exceed the perturbative limit. 
However, in that regime, a light dark $Z$ mass cannot be obtained, leading to no deviation from the Standard Model 
predictions for the $B\to K\nu\bar{\nu}$ and $B\to K^*\nu\bar{\nu}$ decays.

\begin{figure*}[htb]
	\includegraphics[width=0.95\linewidth]{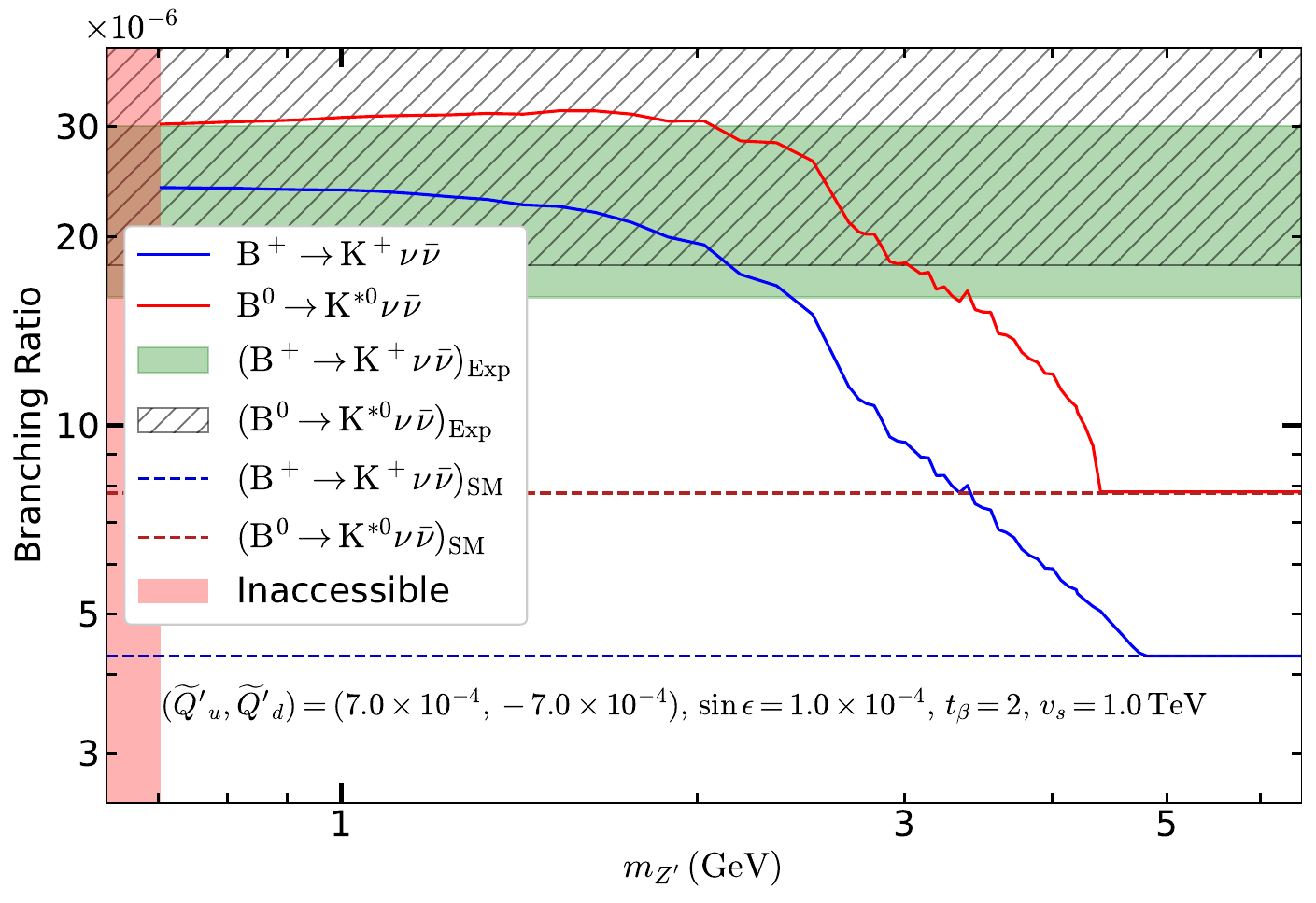}
	\caption{The branching ratios of $B\to K^{(*)}\nu\bar{\nu}$ decays as a function of dark $Z$ mass $m_{Z^{'}}$ for representative values of the input variables. Belle II measurement \cite{Belle-II:2023esi} of ${\rm Br}(B^+\to K^+\nu\bar{\nu})$ is displayed by the green band, while the exclusion region by ${\rm Br}(B^0\to K^{*0}\nu\bar{\nu})$ from Belle data \cite{Belle:2017oht} is hatched. The SM values of both decays are also indicated in the figure. The red shaded region is excluded due to inaccessible $Z^{'}$ masses.}
	\label{fig:brmapexample1}
\end{figure*}

Another and perhaps the most important observation obtained from our analysis is that when performing the $q^2$ integral encountering the kinematic situation where $q^2$ equals the mass of the dark $Z$ allows the modification of the dark $Z$ propagator in the relevant diagrams to incorporate the width effects. This modification enables deviations above the Standard Model values and, in particular, allows the explanation of the Belle data for the $B\to K\nu\bar{\nu}$ decay.

 The variation of the branching ratio of the $B\to K\nu\bar{\nu}$ decay with respect to the mass of the dark $Z$ is presented in Fig.~\ref{fig:brmapexample1}. There is a lower limit on the mass of the dark $Z$, which can be determined by setting $m_X$ to zero once a set of input values are chosen for the other parameters. For example, for the input values given in Fig.~\ref{fig:brmapexample1}, the minimum value of mass that the dark $Z$ can have is close to $70$ MeV. Then, such a region with inaccessible dark $Z$ masses is shown as the red shaded color. 
 
In this calculation, it is also necessary to determine the decay width of the dark $Z$, $\Gamma_{Z^{'} }$, by incorporating new channels as the mass varies. The effects of these new channels can be seen in Fig.~\ref{fig:brmapexample1}. On the other hand, it should be noted that the results presented here do not exhibit significant sensitivity to the parameters $\sin\epsilon$ or $\tan\beta$. The numerical value of $\Gamma_{Z^{'}}$, which is vital for deviations from the SM, depends on various model parameters. For example, for the inputs given in Fig.~\ref{fig:brmapexample1}, $\Gamma_{Z^{'}}\simeq 0.1$ keV  when $m_{Z^{'}}=4.5$ GeV.

Deviations from the Standard Model start to appear when the dark $Z$ mass is around 5 GeV, and the results become compatible with Belle II data when the mass is approximately 2 GeV. However, when the mass drops below 1 GeV, the dependence on mass disappears. Similar behavior is observed for the $B\to K^*\nu\bar{\nu}$ decay. Although the mass dependence occurs at a slightly lower value compared to the $B\to K\nu\bar{\nu}$ decay, the branching ratio exceeds the experimental upper limit at around 3 GeV. Consequently, under the chosen input values, there is no common parameter space that simultaneously satisfies both results.

\begin{figure}[htb]
	$\begin{array}{cc}
		\includegraphics[width=0.5\linewidth,height=6cm]{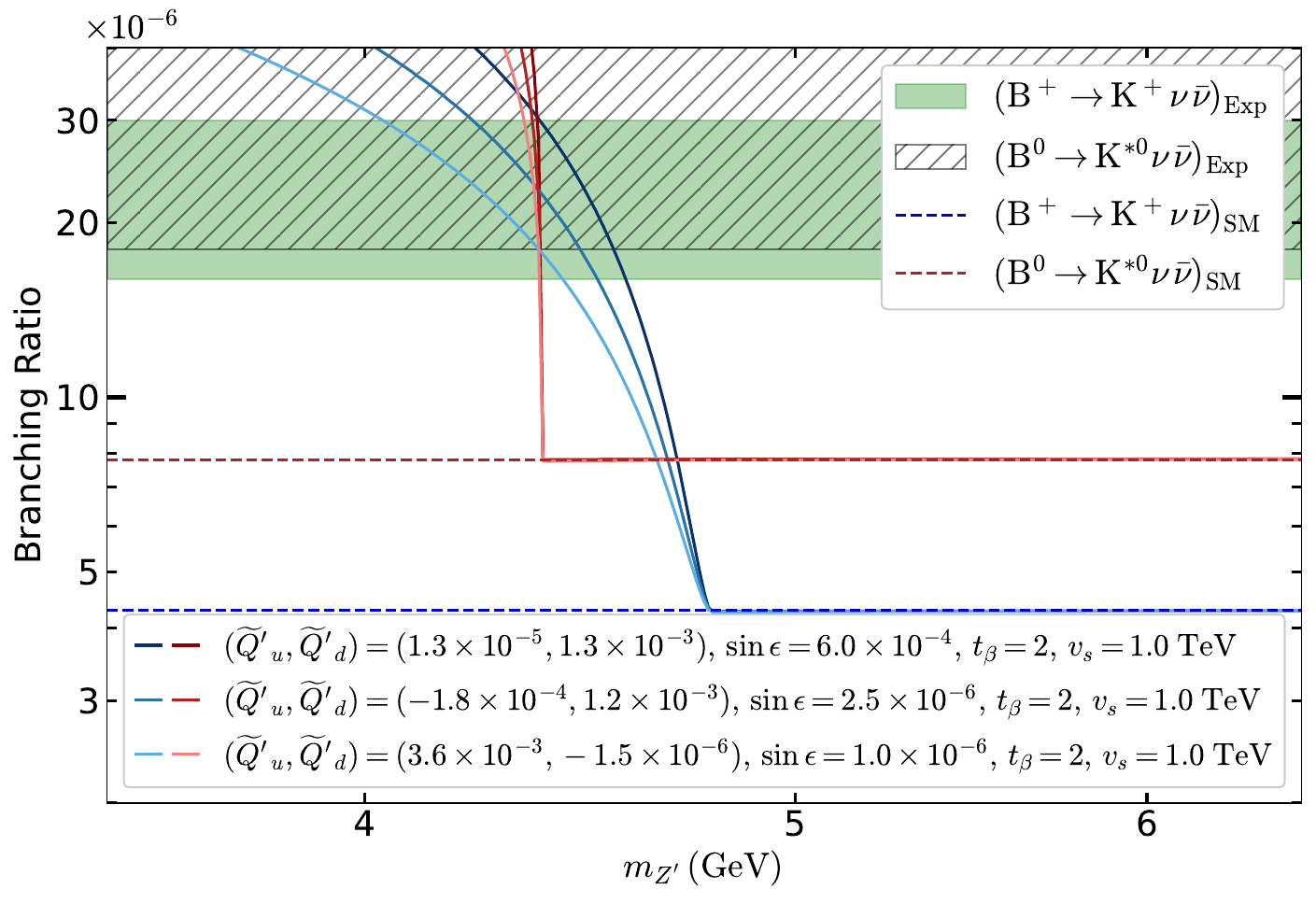} 
		\label{fig:subim1}
		&
		\includegraphics[width=0.5\linewidth,height=6cm]{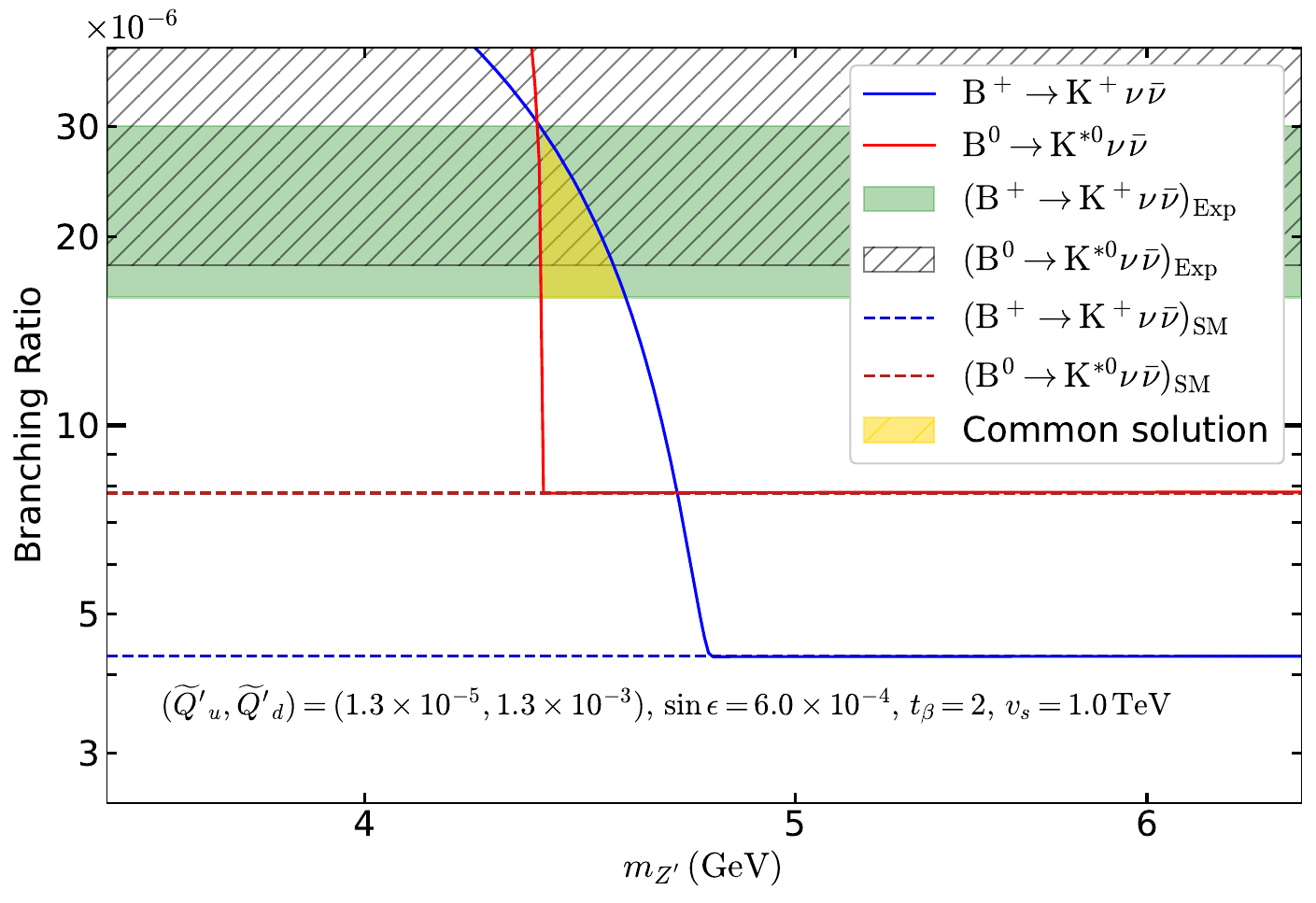}
		\label{fig:subim2}
	\end{array}$
	\vskip -0.3cm
	\caption{The branching ratios of $B\to K^{(*)}\nu\bar{\nu}$ decays as a function of dark $Z$ mass $m_{Z^{'}}$ for three different values of the input variables (left panel) and for a scenario with a maximized allowed region (right panel).  The Belle II measurement \cite{Belle-II:2023esi} of ${\rm Br}(B^+\to K^+\nu\bar{\nu})$ is displayed in the green band, while the hatched region as the exclusion region determined by the upper bound of ${\rm Br}(B^0\to K^{*0}\nu\bar{\nu})$ from Belle data \cite{Belle:2017oht}. The SM values of both decays are also indicated in the figures. In the right panel, a maximized common solution region is highlighted.}
	\label{fig:brmapexample2}
\end{figure}

In Fig.~\ref{fig:brmapexample2}, we investigated whether a common solution region exists. The left panel shows that a common solution is only possible in the 4–5 GeV range and how the allowed region shrinks as the parameters vary. In the right panel, the parameter values that yield the largest possible region are given, and the common region has been scanned. As can be seen, the common solution region is quite small, but it is still possible to explain both sets of data. There is one important feature that needs to be noted here. Even if no excess events can be reported for  $B\to K^*\nu\bar{\nu}$ and the current upper limit decreases further, potentially even reaching the SM value, the outcomes we have derived here will remain unaffected. We note that the decay width of $Z^{'}$ becomes around 1 keV for the inputs in the right panel of   Fig.~\ref{fig:brmapexample2}. Hence, in the laboratory frame of Belle II detector, the typical decay length of the $Z'$ boson is very subdued so that it decays within the detector. 

\begin{figure}[htb]
	\centering
	\includegraphics[width=0.95\linewidth]{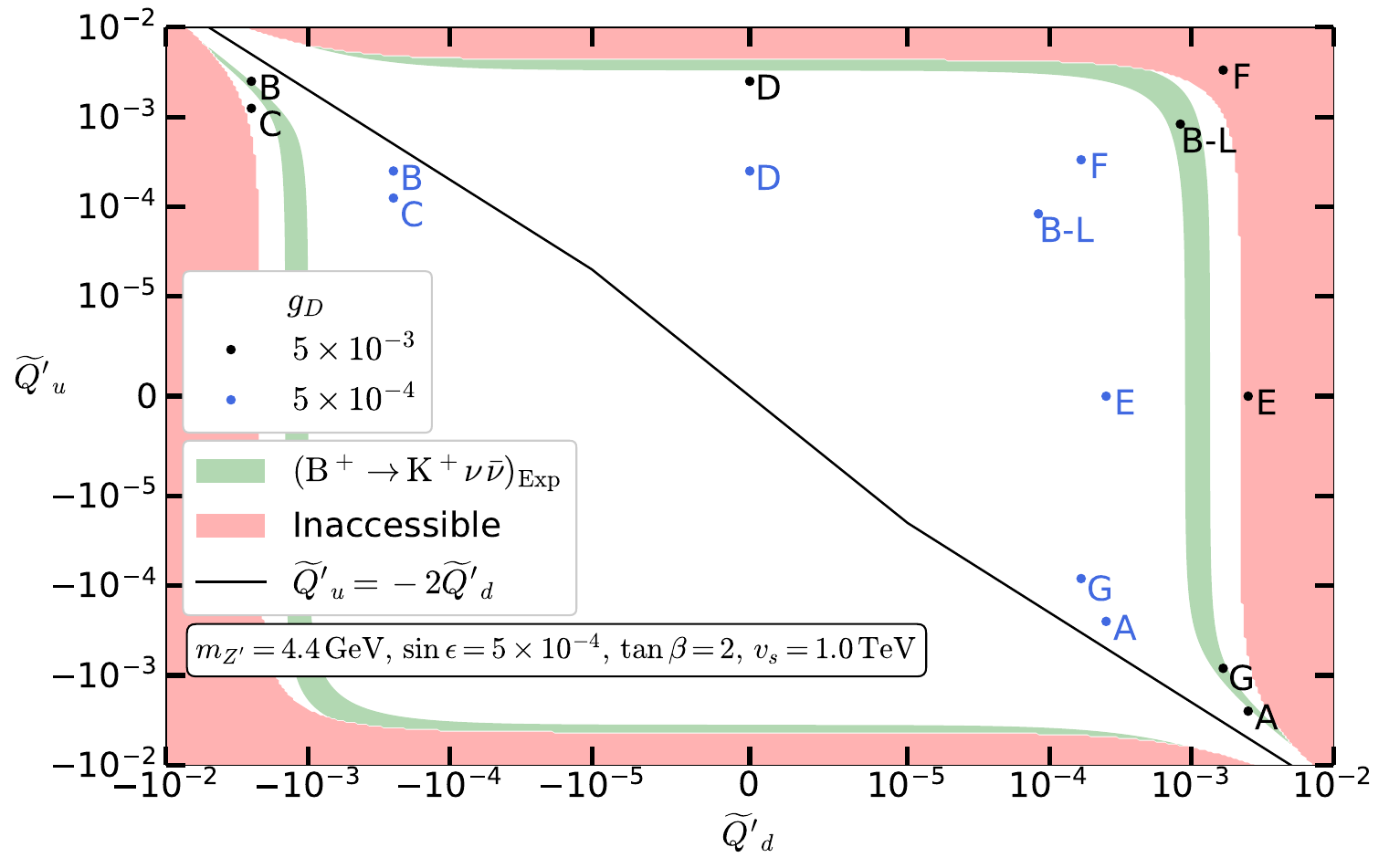}
	\caption{The Belle II measurement \cite{Belle-II:2023esi} of ${\rm Br}(B^+\to K^+\nu\bar{\nu})$ (thin green shaded area)  is shown in the $(\widetilde{Q}_d^{'},\widetilde{Q}_u^{'})$ plane for the favored dark $Z$ mass. Points representing the models listed in Table~\ref{ChargeTable} are marked for two different $g_D$ values. The Belle data \cite{Belle:2017oht} for ${\rm Br}(B^0\to K^{*0}\nu\bar{\nu})$ are not shown since it remains under the red shaded area, representing the inaccessible region by dark $Z$ mass considerations. Additionally, the line representing $\widetilde{Q}_u^{'}=-2\widetilde{Q}_d^{'}$ is shown in black and it marks the cases where the neutral flavor conserving criterion is violated \cite{Campos:2017dgc}.}
	\label{fig:brmapexample3}
\end{figure}

If we want to visualize the allowed regions from the ${\rm Br}(B\to K\nu\bar{\nu})$ measurement along with the inaccessible regions from the dark $Z$ mass considerations in the $(\widetilde{Q}_d^{'},\widetilde{Q}_u^{'})$ plane, we obtain Fig.~\ref{fig:brmapexample3}. The ${\rm Br}(B\to K^*\nu\bar{\nu})$ data are not shown since they remain under the unphysical region indicated by the red shaded color. As seen in the figure, for the selected input values, there exists a narrow green-shaded region favored by ${\rm Br}(B\to K\nu\bar{\nu})$, while the excluded region from ${\rm Br}(B\to K^*\nu\bar{\nu})$ does not overlap with it. To illustrate the new physics models selected as representative points in such a parameter space, we need to specify a value for $g_D$. As an example, we consider $g_D = 5\times 10^{-4}$ and $g_D = 5\times 10^{-3}$. Particularly, for $g_D = 5\times 10^{-3}$, some models lie within the green region, whereas for smaller values of $g_D$, all models can be excluded. We believe that the ${\rm Br}(B\to K^{(*)}\nu\bar{\nu})$ data significantly constrain the new physics parameter space and can play a decisive role in distinguishing between possible scenarios. We like to note that, while one might identify the point  $( \widetilde{Q}_d^{'} , \widetilde{Q}_u^{'} )=(0,0)$ in Fig.~\ref{fig:brmapexample3} as representing the SM, this interpretation is not entirely accurate. Although this point corresponds to $g_D=0$ (and/or zero dark charges), the presence of nonzero mixing (e.g., $\sin\epsilon\ne 0$) means that the $Z$ contribution still differs from that of the SM $Z$ boson. To avoid potential misinterpretation, we deliberately chose not to designate any specific point as the Standard Model. Furthermore, given that the SM has been demonstrated to be insufficient to account for the Belle~II measurements of ${\rm Br}(B^+\to K^+\nu\bar{\nu})$, our primary objective in this plot is to identify which parameter combinations of $(\widetilde{Q}_d^{'}, \widetilde{Q}_u^{'})$ together with the other free parameters of the model and under which specific conditions various representative models could fall within the experimentally favored green shaded region.

\begin{figure}[htb]
	\centering
	\includegraphics[width=0.95\linewidth]{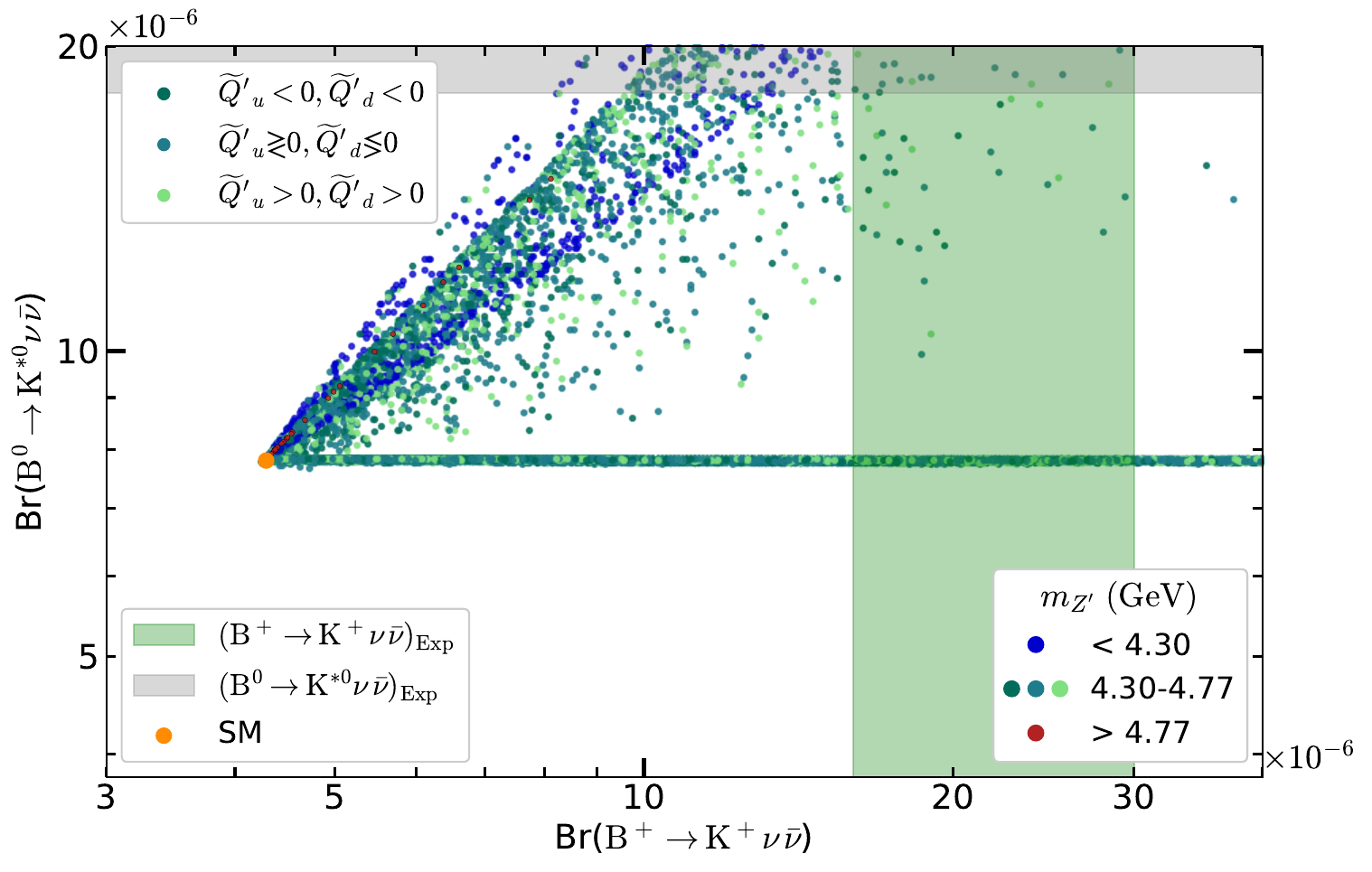}
	\caption{Scatter plot in  the $({\rm Br}(B^+\to K^+\nu\bar{\nu}), {\rm Br}(B\to K^*\nu\bar{\nu}))$ plane for various sign conventions of $(\widetilde{Q}_d^{'},\widetilde{Q}_u^{'})$. The rare red points represent a scenario with $m_{Z^{'}}\ge 4.77$ GeV, while the blue ones are for $m_{Z^{'}}\le 4.3$ GeV. The remaining green dots (varying tones encoded with changing signs of $(\widetilde{Q}_d^{'},\widetilde{Q}_u^{'})$ parameters) represent  4.3 GeV$ < m_{Z^{'}} < 4.77$ GeV.  }
	\label{fig:brmapexample4}
\end{figure}

Finally, we generated a scatter plot in the $B\to K\nu\bar{\nu}$ versus $B\to K^*\nu\bar{\nu}$ plane. The following parameter ranges were considered:  
\begin{eqnarray}
	10^{-3} < \left| \widetilde{Q}_{u,d}^{'}\right| < 10^{3}\,, && \qquad
	 10^{-6} < \sin\epsilon < 10^{-2}\,,  \nonumber\\
	 1 < \tan\beta < 60\,,  && \qquad
     10\, {\rm MeV} < m_{Z^'} < 10\, {\rm GeV}\,.
\end{eqnarray}

Under these conditions, Fig.~\ref{fig:brmapexample4} was obtained, and as can be seen from the figure, neither red nor blue points can enter the allowed region of $B\to K\nu\bar{\nu}$ without going through the excluded region of $B\to K^*\nu\bar{\nu}$. Only the green points satisfy both conditions, indicating that the allowed mass range is 4.3 GeV $< m_{Z^{'}} < 4.77$ GeV since the red (blue) points represent a scenario with $m_{Z^{'}}\ge 4.77$ GeV ($m_{Z^{'}}\le 4.3$ GeV). These results depend particularly on the values and signs of $\widetilde{Q}_u^{'}$ and $\widetilde{Q}_d^{'}$. In the figure, scenarios with both $\widetilde{Q}_d^{'}$ and $\widetilde{Q}_u^{'}$ positive are encoded with light green points while the darker ones represent negative case for both of them. Parameters with opposite signs are indicated by the dark green color. The green points that appear parallel to the $B\to K\nu\bar{\nu}$ axis correspond to the region where the branching ratio of $B\to K^*\nu\bar{\nu}$ is not yet affected by the dark $Z$, while the branching ratio of $B\to K\nu\bar{\nu}$ increases rapidly. This region is also indicated in Fig.~\ref{fig:brmapexample2}. Even though the overall solution space satisfying both constraints is rather small, it is seen that, out of 85,000 points generated under the conditions described above, relatively quite a number of them fall into the desired region.

\section{Conclusion}
\label{conc}
\raggedbottom
The recent Belle II measurement on rare $B$ decay $B^+\to K^+\nu\bar{\nu}$ showed a $2.7\sigma$ deviation from the SM value, which can be seen a fertile ground for searching new physics. The Belle Collaboration additionally has data on another rare $B$ meson channel, namely, $B^0\to K^{*0}\nu\bar{\nu}$, setting an upper bound $1.8\times 10^{-5}$ for the branching ratios of the decay. To test whether a new physics scenario in the form of a two-Higgs-doublet model extended in the gauge sector with an Abelian dark gauge group $U(1)_D$ would account for the discrepancy mentioned above as well as satisfying the exclusion bound for  $B^0\to K^{*0}\nu\bar{\nu}$ decay, an extensive numerical study has been performed. Both processes happen at one-loop level, and their Feynman diagrams can be classified as self-energy, triangle, and box. The dark $Z$ contributes only to self-energy and triangle diagrams. The matrix element of each diagram is computed and expressed in terms of the Passarino-Veltman functions in Appendix \ref{sec:appenA} in the vanishing limit of the masses and momenta of the external particles. We have made extensive checks on our results to make sure that the total matrix element is UV finite. Our analytical findings are further manipulated to carry loop integrals which could be tricky under the limit employed so that a comparison with the results available in the literature. We found full agreement in the Standard Model limit  with the results given in Refs. \cite{Voloshin:1976in,Inami:1980fz,Wang:2023css}.

Let us stress the following point one more time. The first important observation we have made is that  the discrepancy between the Belle II data and the SM prediction for the decay   $B^+\to K^+\nu\bar{\nu}$ would not be elucidated by the existence of a dark $Z$ unless the resonant contributions around  $q^2=m_{Z^{'}}^2$ are taken into account. 
Far from the resonance region, the dark $Z$ contribution becomes much less significant and no deviation would be obtained from the SM values of both channels. 

Even though satisfying either bound individually is rather easy to achieve strictly after the resonant contributions are manifested in the analytical level, finding a common parameter space which is good for both requires some further analysis. Our numerical study showed that, while the exclusion data of  $B^0\to K^{*0}\nu\bar{\nu}$ prefer a heavier dark $Z$ scenario, $B^+\to K^+\nu\bar{\nu}$ measurement would indicate very light dark $Z$ options which can get as heavy as few GeV. The task is to seek  a common mass region which lies in the overlapped potion of the parameter space. As shown in the figures, for reasonable values of the other parameters of the model ($\widetilde{Q}_u^{'}, \widetilde{Q}_d^{'},\sin\epsilon, \tan\beta$), a dark $Z$ with a 4-5 GeV mass would explain both. Even though the sensitivity of our results is not affected muchby the parameters $\sin\epsilon$ or $\tan\beta$, the values of the other two parameters $\widetilde{Q}_u^{'}$ and $\widetilde{Q}_d^{'}$ are playing more essential role. On the other hand, our scatter plot would prove that the overlapping region for the mass of the dark $Z$ does not shift much and remains in the $4$-$5$ GeV range. Another important feature seen from the figures is that, even if no excess can be reported for the  ${\rm Br}(B^0\to K^{*0}\nu\bar{\nu})$ channel and the existing upper limit value decreases further, even dropping to the SM value, the results we have obtained here will not be affected by this. Current bounds already pin down  the allowed range of the dark $Z$ mass in particular and future improvements to reduce the errors further  in the $B^+\to K^+\nu\bar{\nu}$ decay will tighten up the allowed region more and will determine whether a common solution would still be possible or not. 
\acknowledgments
This study was supported by Scientific and Technological Research Council of Turkey (TUBITAK) under the Grant Number 124F237. The authors thank TUBITAK for its support.

\appendix
\section{Explicit Expressions of the Coefficients for $C_{LL}$ and $C_{LR}$}
\label{sec:appenA}
In this Appendix we present the explicit expressions of the coefficients $C_{LL}$ and $C_{LR}$ computed from the triangle, self-energy and box diagrams.\footnote{After computing each diagram, we checked the SM limits, whenever relevant, and found agreement with the results available in the literature. Additionally, the $UV$ finiteness of the overall result has been confirmed after especially the factor of 1/2 taken into account Ref. \cite{Voloshin:1976in}.} Before presenting these expressions for customary we introduce following short-handed notations:
\begin{align*}
	x &= m_t^2/m_W^2, \, y = m_e^2/m_W^2,  \\
	\left[\begin{array}{c}
	C_L^\nu\\
	{C'}_L^\nu
\end{array} \right]  & = \frac{1}{4}  
\left[\begin{array}{c}
	\sin \xi \\
	\cos \xi
\end{array} \right]  \left(2 \tan \epsilon  \sec \theta _W-\frac3{e} (\widetilde{Q}_d^'+\widetilde{Q}_u^') \sec \epsilon \right)+ \left[\begin{array}{c}
\cos \xi \\
-\sin \xi
\end{array} \right]  \csc 2 \theta _W, \\
	\left[\begin{array}{c}
	C_R^\nu\\
	{C'}_R^\nu
\end{array} \right]  &= 	 -\frac{1}{2e}  \left[\begin{array}{c}
\sin \xi \\
\cos \xi
\end{array} \right]  (2 \widetilde{Q}_d^'+\widetilde{Q}_u^') \sec \epsilon , \\	
	\left[\begin{array}{c}
			C_L^u\\
			C_L^d
		\end{array} \right]  & = \frac{1}{12}  \left[\sin \xi  \left(\frac3{e}  (\widetilde{Q}_d^'+\widetilde{Q}_u^') \sec \epsilon -2 \tan \epsilon  \sec \theta _W\right)+ \cos \xi  \left(\pm 6 \cot \theta _W - 2 \tan \theta _W\right)\right], \\
\left[\begin{array}{c}
	C_R^u\\
	C_R^d
\end{array} \right]  & = \frac{1}{6}  \left(\frac3{e} \left[\begin{array}{c}
\widetilde{Q}_u^'\\
	\widetilde{Q}_d^'
\end{array} \right] \sin \xi  \sec \epsilon +\left[\begin{array}{c}
-4\\
2
\end{array} \right] \left(\cos \xi  \tan \theta _W +  \sin \xi  \tan \epsilon  \sec \theta _W\right) \right),  \\
\left[\begin{array}{c}
	{C'}_L^u\\
	{C'}_L^d
\end{array} \right]  & = \frac{1}{12}  \left[\cos \xi  \left(\frac3{e} (\widetilde{Q}_d^'+\widetilde{Q}_u^') \sec \epsilon -2 \tan \epsilon  \sec \theta _W\right)+ \sin \xi  \left(\mp 6 \cot \theta _W + 2 \tan \theta _W\right)\right], \\
\left[\begin{array}{c}
	{C'}_R^u\\
	{C'}_R^d
\end{array} \right]  & = \frac{1}{6}  \left(\frac3{e} \left[\begin{array}{c}
	\widetilde{Q}_u^'\\
	\widetilde{Q}_d^'
\end{array} \right] \cos \xi  \sec \epsilon +\left[\begin{array}{c}
	-4\\
	2
\end{array} \right] \left(-\sin \xi  \tan \theta _W +  \cos \xi  \tan \epsilon  \sec \theta _W\right) \right),  \\
%
			\left[\begin{array}{c}
		G^{WGZ}\\
		G^{WGZ^\prime }
	\end{array} \right]  & =   
	\sin \beta \csc \theta _W \left( \left[\begin{array}{c}
		\sin \xi \\
		-\cos \xi
	\end{array} \right]  \left(\frac1{e} (\widetilde{Q}_d^'-\widetilde{Q}_u^') \sec \epsilon +2 \tan \epsilon  \sec \theta _W \right)+2 \left[\begin{array}{c}
		\cos \xi \\
		\sin \xi
	\end{array} \right]\tan \theta _W \right)  ,\\
		\left[\begin{array}{c}
		G^{GGZ}\\
		G^{GGZ^\prime }
	\end{array} \right]  & = \frac{1}{4}  
	\left[\begin{array}{c}
		\sin \xi \\
		\cos \xi
	\end{array} \right]  \left(\frac1{e} (\widetilde{Q}_d^'-\widetilde{Q}_u^') \sec \epsilon +2 \tan \epsilon  \sec \theta _W \right)+\frac12  \left[\begin{array}{c}
		\cos \xi \\
		-\sin \xi
	\end{array} \right] \left(1-\cot ^2\theta _W\right) \,.
\end{align*}  

\subsection{Triangle Z Diagrams }
\label{TriangleZD} 

\begin{figure}[h!]
	\centering
	\includegraphics[scale=1.2]{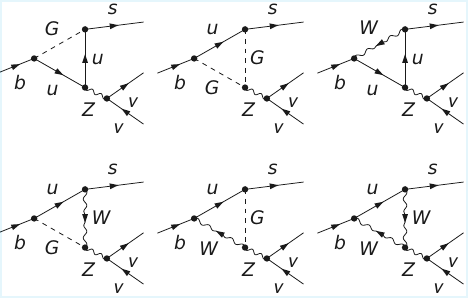}
	\caption{Triangle $Z$ diagrams relevant for $b \rightarrow s\nu \bar\nu$. Only one neutrino generation is shown.}
			\label{fig:trigZ}
\end{figure}

The total result obtained from the triangle diagrams in Fig.~\ref{fig:trigZ} for the $Z$ boson exchange is 

\begin{align}
	{\cal M}_Z^{\text{Tri}}= \frac{V_{tb} V_{ts}^* G_F \, \alpha _{em} \, m_W^2}{2\sqrt{2} \pi (m_Z^2-q^2)} \left(C_{LL}^{TZ} {\cal O}_{LL} + C_{LR}^{TZ} {\cal O}_{LR}\right)
\end{align}
where 
\begin{align}
	C_{LL}^{TZ} =& C_L^\nu \Bigg \{ \Bigg[
	+ C_0(0,0,0;  m_W^2, m_W^2, m_W^2) 4 m_W^2 \cos \xi \cot \theta_W \nonumber \\
	& - C_0(0,0,0;  m_W^2, m_t^2, m_t^2) 2 m_t^2   \left(  C_L^u x + 2 C_R^u   \right)\nonumber \\
	& - C_0(0,0,0;  m_t^2, m_W^2, m_W^2) \frac{2 m_t^2}{\pi} \left(G^{WGZ} \sin \theta_W + 2 \pi  \cos \xi \cot \theta_W   \right) \nonumber \\
	& - C_{00}(0,0,0;  m_W^2, m_W^2, m_W^2) 16 \cos \xi \cot \theta_W \nonumber \\
	& + C_{00}(0,0,0;  m_W^2, m_t^2, m_t^2) 4 ( 2 C_L^u + C_R^u x)\nonumber \\
	& + C_{00}(0,0,0;  m_t^2, m_W^2, m_W^2) 4 \left( G^{GGZ} x - 2 \cos \xi \cot \theta_W \right)\nonumber \\
	& + 4 \cos \xi \cot \theta_W - (4 C_L^u + C_R^u x)
	\Bigg] - (m_t \rightarrow 0)
	\Bigg \}
\end{align}

and
\begin{align}
	C_{LR}^{TZ} =& C_R^\nu\Bigg \{ \Bigg[
	+ C_0(0,0,0;  m_W^2, m_W^2, m_W^2) 4 m_W^2 \cos \xi \cot \theta_W \nonumber \\ 
	& - C_0(0,0,0;  m_W^2, m_t^2, m_t^2) 2 m_t^2   \left(  C_L^u x + 2 C_R^u   \right)\nonumber \\
	& - C_0(0,0,0;  m_t^2, m_W^2, m_W^2) \frac{2 m_t^2}{\pi} \left(G^{WGZ} \sin \theta_W  + 2 \pi  \cos \xi \cot \theta_W  \right)\nonumber \\
	& - C_{00}(0,0,0;  m_W^2, m_W^2, m_W^2) 16 \cos \xi \cot \theta_W \nonumber \\
	& + C_{00}(0,0,0;  m_W^2, m_t^2, m_t^2) 4 ( 2 C_L^u + C_R^u x)\nonumber \\
	& + C_{00}(0,0,0;  m_t^2, m_W^2, m_W^2) 4 \left( G^{GGZ} x - 2 \cos \xi \cot \theta_W \right)\nonumber \\
	& + 4 \cos \xi \cot \theta_W -(4 C_L^u + C_R^u x)
	\Bigg] - (m_t \rightarrow 0)
	\Bigg \}\,.
\end{align}
 

\subsection{Triangle $Z^{'}$ Diagrams }
\label{TriangleZpD} 

\begin{figure}[h!]
	\centering
	\includegraphics[scale=1.2]{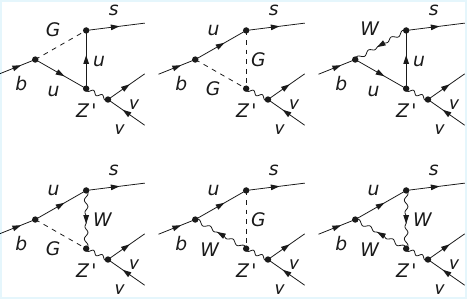}
		\caption{Triangle $Z^{'}$ diagrams relevant for $b \rightarrow s\nu \bar\nu$. Only one neutrino generation is shown.}
		\label{fig:trigAp}
\end{figure}

The matrix element obtained from the triangle diagrams in Fig.~\ref{fig:trigAp} the dark $Z$ exchange is: 
\begin{align}
	{\cal M}_{Z^\prime }^{\text{Tri}}= \frac{V_{tb} V_{ts}^* G_F \, \alpha _{em} \, m_W^2}{2\sqrt{2} \pi (m_{Z^\prime}^2 - q^2) }  \left(C_{LL}^{TZ^\prime} {\cal O}_{LL} + C_{LR}^{TZ^\prime } {\cal O}_{LR}\right)
\end{align}
where 
\begin{align}
	C_{LL}^{TZ^\prime } =& {C'}_L^\nu \Bigg \{ \Bigg[
	-C_0(0,0,0;  m_W^2, m_W^2, m_W^2) 4 m_W^2 \sin \xi \cot \theta_W \nonumber \\
	& -2 C_0(0,0,0;  m_W^2, m_t^2, m_t^2) 2 m_t^2 (2 {C'}_R^u + {C'}_L^u x)  \nonumber \\
	&+ C_0(0,0,0;  m_t^2, m_W^2, m_W^2) \frac{2 m_t^2}{\pi} \left(G^{WGZ^\prime } \sin \theta_W  + 2 \pi  \sin \xi \cot \theta_W  \right) \nonumber \\
	& + C_{00}(0,0,0;  m_W^2, m_W^2, m_W^2) 16 \sin \xi \cot \theta_W \nonumber \\
	& + C_{00}(0,0,0;  m_W^2, m_t^2, m_t^2) 4 ( 2 {C'}_L^u + {C'}_R^u x)\nonumber \\
	& + C_{00}(0,0,0;  m_t^2, m_W^2, m_W^2) 4 \left( G^{GGZ^\prime } x - 2 \sin \xi \cot \theta_W \right)\nonumber \\
	& - 4 \sin \xi \cot \theta_W -(4 {C'}_L^u + {C'}_R^u x)
	\Bigg] - (m_t \rightarrow 0)
	\Bigg \}
\end{align}
and 
\begin{align}
	C_{LR}^{TZ^\prime } =&  {C'}_R^\nu \Bigg \{ \Bigg[
	-C_0(0,0,0;  m_W^2, m_W^2, m_W^2) 4 m_W^2 \sin \xi \cot \theta_W \nonumber \\
	& -2 C_0(0,0,0;  m_W^2, m_t^2, m_t^2) 2 m_t^2 (2 {C'}_R^u + {C'}_L^u x)  \nonumber \\
	&+ C_0(0,0,0;  m_t^2, m_W^2, m_W^2) \frac{2 m_t^2}{\pi} \left(G^{WGZ^\prime } \sin \theta_W  + 2 \pi  \sin \xi \cot \theta_W  \right) \nonumber \\
	& + C_{00}(0,0,0;  m_W^2, m_W^2, m_W^2) 16 \sin \xi \cot \theta_W \nonumber \\
	& + C_{00}(0,0,0;  m_W^2, m_t^2, m_t^2) 4 ( 2 {C'}_L^u + {C'}_R^u x)\nonumber \\
	& + C_{00}(0,0,0;  m_t^2, m_W^2, m_W^2) 4 \left( G^{GGZ^\prime } x - 2 \sin \xi \cot \theta_W \right)\nonumber \\
	& - 4 \sin \xi \cot \theta_W -(4 {C'}_L^u + {C'}_R^u x)
	\Bigg] - (m_t \rightarrow 0)
	\Bigg \}
\end{align}

\subsection{Self-energy $Z$ and  $Z^{'}$ Diagrams }
\label{SelfenergyZZpD}

Finally, we present the self-energy diagrams due to the $Z$ and dark $Z^{'} $ exchange.

\subsubsection{Self-energy $Z$: }

\begin{figure}[h!]
	\centering
	\includegraphics[scale=0.9]{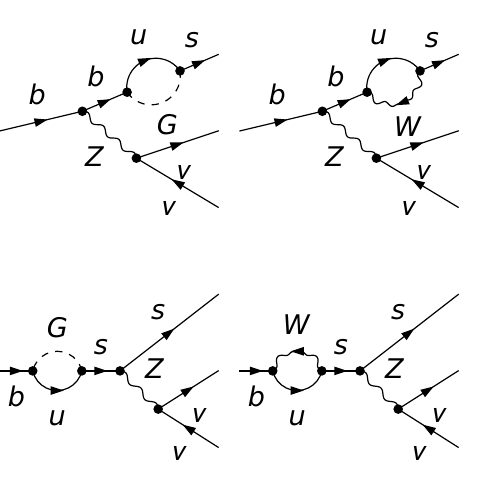}
	\caption{Self-energy $Z$ diagrams relevant for $b \rightarrow s\nu \bar\nu$. Only one neutrino generation is shown.}	
	\label{fig:selfZ}
\end{figure}

Using the Feynman diagrams in Fig.~\ref{fig:selfZ}, the result for the self energy diagram contributions due to the $Z$ boson exchange becomes 

\begin{align}
	{\cal M}_Z^{\text{Self}}= \frac{V_{tb} V_{ts}^* G_F \, \alpha _{em} \, m_W^2}{2\sqrt{2} \pi (m_Z^2-q^2)}  \left(C_{LL}^{SZ} {\cal O}_{LL} + C_{LR}^{SZ} {\cal O}_{LR}\right)
\end{align}
where
\begin{align}
	C_{LL}^{SZ} =& 2 C_L^\nu C_L^d  \Big \{  \Big[
	B_1(0; m_t^2, m_W^2) (2+x ) + 1  \Big] - (m_t \rightarrow 0)
	\Big \}
\end{align}
and 
\begin{align}
	C_{LR}^{SZ} =& 2 C_R^\nu C_L^d   \Big \{ \Big[
	B_1(0; m_t^2, m_W^2) (2+x ) + 1  \Big] - (m_t \rightarrow 0)
	\Big \}
\end{align}

\subsubsection{Self-energy $Z^{'} $: }

\begin{figure}[h!]
	\centering
	\includegraphics[scale=0.85]{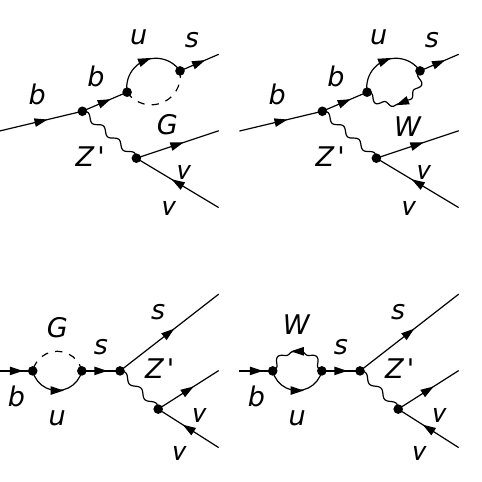}
		\caption{Self-energy $Z^{'} $ diagrams relevant for $b \rightarrow s\nu \bar\nu$. Only one neutrino generation is shown.}	
	\label{fig:selfAp}
\end{figure}

The result for the self-energy diagram contributions due to the dark $Z^{'} $ exchange, computed from the diagrams in Fig.~\ref{fig:selfAp} is 

\begin{align}
	{\cal M}_{Z^{'} }^{\text{Self}}= \frac{V_{tb} V_{ts}^* G_F \, \alpha _{em} \, m_W^2}{2\sqrt{2} \pi (m_{Z^{'}}^2 - q^2)}\left(C_{LL}^{SZ^\prime } {\cal O}_{LL} + C_{LR}^{SZ^\prime } {\cal O}_{LR} \right)
\end{align}
where
\begin{align}
	C_{LL}^{SZ^\prime } =& 2 {C'}_L^\nu {C'}_L^d  \Big \{ \Big[
	B_1(0; m_t^2, m_W^2) (2+x) + 1 \Big] - (m_t \rightarrow 0)
	\Big \}
\end{align}
and 
\begin{align}
	C_{LR}^{SZ^\prime} =& 2 {C'}_R^\nu {C'}_L^d  \Big \{ \Big[
	B_1(0; m_t^2, m_W^2) (2+x) + 1 \Big] - (m_t \rightarrow 0)
	\Big \}
\end{align}

\subsection{Box Diagrams }

\begin{figure}[h!]
	\centering
	\includegraphics[scale=0.85]{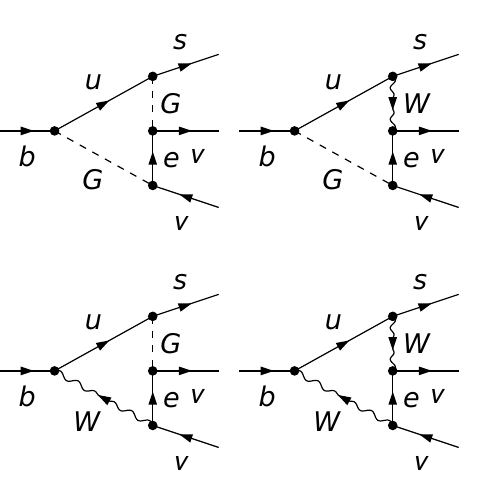}
		\caption{Box diagrams relevant for $b \rightarrow s\nu \bar\nu$. Only one neutrino generation is shown.}	
	\label{fig:box}
\end{figure}

Finally, the box diagram amplitude from Fig.~\ref{fig:box} is
\begin{align}
	{\cal M}^{\text{Box}}= \frac{V_{tb} V_{ts}^* \, G_F \, \alpha _{em} \, m_W^2}{2\sqrt{2} \pi}\   C_{LL}^{Box} {\cal O}_{LL} 
\end{align}
where
\begin{align}
	C_{LL}^{Box} =&    \Big[
	-2 D_0(0, 0, 0, 0,0,0; m_t^2, m_W^2, m_e^2, m_W^2) \; m_W^2 x y \nonumber\\
	& +  D_{00}(0, 0, 0, 0,0,0; m_t^2, m_W^2, m_e^2, m_W^2) (16+xy)\Big] -(m_t \rightarrow 0)
\end{align}

It should be noted that the dark $Z$ do not contribute to the box diagrams. The one-loop integral expressions $B_i$, $C_0$, $C_{00}$, $D_0$, and $D_{00}$ are the standard Passarino-Veltman functions and they are defined by using the convention in Ref.  \cite{Hahn:1998yk}.    

\bibliography{mycevensV2}

\end{document}